\documentclass[prb,twocolumn,showpacs,amsmath,amssymb,citeautoscript]{revtex4}

\usepackage{amssymb}
\usepackage{amsmath}
\usepackage{colordvi}
\usepackage[colorlinks]{hyperref}
\usepackage{amsthm}
\usepackage{subeqnarray}
\usepackage{cases}
\usepackage{enumerate}
\usepackage{graphicx}
\usepackage{epsfig}
\usepackage{epstopdf}
\usepackage{subfigure}
\usepackage{color}
\usepackage{multirow}
\usepackage{ulem}

\def\avg(#1){\langle#1\rangle}

\def\be{\begin{equation}}
\def\ee{\end{equation}}
\def\bea{\begin{eqnarray}}
\def\eea{\end{eqnarray}}

\def\nn{\nonumber}

\begin{document}
\title{Tunneling frustration induced peculiar supersolid phases in the extended Bose-Hubbard model}
\author{Shao-Jun Dong}
\author{Wenyuan Liu}
\author{Xiang-Fa Zhou}
\email{xfzhou@ustc.edu.cn}
\author{Guang-Can Guo}
\author{Zheng-Wei Zhou}
\author{Yong-Jian Han}
\email{smhan@ustc.edu.cn}
\author{Lixin He}
\email{helx@ustc.edu.cn}
\affiliation{Key Laboratory of Quantum Information, Chinese Academy of Sciences, University of Science and Technology of China, Hefei, 230026, China}
\affiliation{Synergetic Innovation Center of Quantum Information
and Quantum Physics, University of Science and Technology of China, Hefei, 230026, China}
\date{\today}

\begin{abstract}
{By using a state of art tensor network state method,
we study the ground-state phase diagram of an extended Bose-Hubbard model on
the square lattice with frustrated next-nearest neighboring tunneling.
In the hardcore limit, tunneling frustration stabilizes a peculiar half supersolid (HSS) phase
with one sublattice being superfluid and the other sublattice being Mott Insulator away from half filling.
In the softcore case, the model shows very rich phase diagrams above half filling,
including three different types of supersolid phases depending on the interaction parameters.
The considered model provides a promising route to experimentally
search for novel stable supersolid state induced by frustrated tunneling in below half filling region
with dipolar atoms or molecules.
}
\end{abstract}
\pacs{ 03.75.Nt,67.80.kb,67.85.-d}
\maketitle

\section{Introduction}

The investigation of frustrated systems \cite{diep2013frustrated} has received sustained attention in the past years due to its intimate association with different branches of condensed matter physics. The competition in frustrated systems usually greatly enriches the phase diagrams and leads to various exotic quantum phases, such as, quantum spin liquids \cite{anderson1973resonating,balents2010spin}, magnetic domain patterns \cite{wannier1950antiferromagnetism}, and high-Tc superconductivity \cite{1987Anderson,lee2006doping}.
Recently, the experimental success of mimic various theoretical models using ultracold atoms \cite{fisher1989boson,jaksch1998cold,lewenstein2007ultracold,greiner2002quantum} provides a new platform to study the physics of the frustrated systems in a highly controllable manner. For instance,
the frustrated Bose-Hubbard model can emerge when rotating optical lattices in the tight-binding limit,
or with the help of synthetic gauge fields imposed by lasers \cite{dalibard2011colloquium}.
These frustrated systems simulated by the neutral particles \cite{moller2010condensed,huber2010bose,powell2011bogoliubov,inglis2011mott,
dhar2012bose,you2012superfluidity,chan2012evidence,dhar2013chiral,PhysRevB.92.195149,
PhysRevB.93.144508,PhysRevA.81.021602,PhysRevA.85.013632} offer great opportunities to explore new quantum phases.

Experimental searching for the supersolid (SS) phases  is a long-standing challenge and has also received great attention recently.
SS phases are characterized by the simultaneous appearance of crystalline and superfluid orders. Although its existence in $^4$He was
predicted a long time ago,~\cite{andreev1969quantum,chester1970speculations,leggett1970can}
experimental observation the SS remains elusive.\cite{kim2012absence}
Ultracold atoms offer another promising way, it is widely
believed that SS can be stabilized in the extended Bose-Hubbard (EBH) model which can be experimentally
realized by loading dipolar bosons \cite{lahaye2009physics,griesmaier2005bose,lu2011strongly,aikawa2012bose} into an optical lattice.
It has been shown that a stable SS phase can exist in the geometrically frustrated triangular lattice \cite{PhysRevLett.95.237204,PhysRevLett.95.127207,PhysRevLett.95.127206}.
However, in the square lattice, the SS phase becomes fragile resulting in the phase separation (PS) in the hardcore limit \cite{PhysRevLett.74.2527,PhysRevLett.84.1599,PhysRevLett.94.207202,capogrosso2010quantum,
ohgoe2012ground,schmidt2008supersolid,chen2008supersolidity}, especially when filling factor $\rho<1/2$, where the SS phase is missing or only takes a very narrow parameter range irrespective of anisotropic hoppings and interactions \cite{chan2010supersolid,ohgoe2012quantum,ying2013phase}.
How the tunneling frustration change the SS physics remains largely unexplored, because the standard quantum Monte-Carlo method
may suffer from the sign problem.

In this work, we investigate an EBH model with frustrated next-nearest-neighbor (NNN) tunneling on square lattices.
We calculate the ground states of the model via a state of art tensor network state (TNS) method.
We identify a peculiar half SS (HSS) phase that is stable in a large parameter space, where superfluid (SF) and Mott insulator (MI) are simultaneously supported within different sublattice away from half filling due to the presence of NNN hopping and NN repulsion $V$. Increasing the NNN tunneling can enlarge the portion of the HSS phase in the phase diagram, which appears even with very small filling factors. This is very different from the non-frustrated case, where stable SS is always missing in this limit. In the softcore case, the model shows very rich phase diagrams for $\rho>$1/2, including three different types of SS states depending on the relative strength of different interaction parameters.

\section{Model and Methods}

The Hamiltonian considered in the paper reads,
\bea
H &=& -t_1 \sum_{\avg(i,j)}(b_i^{\dag}b_j + h.c.) - t_2 \sum_{\avg(\avg(i,k))} (b_i^{\dag}b_k +
h.c.) \nn \\
&& +\frac{U}{2}\sum_i n_i(n_i-1)+V\sum_{\langle i,j\rangle} n_in_j,
\eea
where $\sum_{\langle i,j\rangle}$ and $\sum_{\langle\langle i,j\rangle\rangle}$ denote the summation of the nearest-neighbor (NN) and the next-nearest-neighbor (NNN) sites with hopping amplitude $t_1$ and $t_2$ respectively, $n_i=b_i^\dag b_i$ is the number operator for boson. $U$ is the on-site interaction energy, $V$ denotes the interaction between NN site. Without loss of generality, we assume $t_1=1$. When $t_2 < 0$, the model shows frustration effects.

The non-frustrated EBH model has been extensively studied using both theoretical and numerical methods
\cite{PhysRevLett.74.2527,PhysRevLett.84.1599,PhysRevLett.94.207202}.
It has also been shown that in triangular lattices \cite{PhysRevLett.95.237204,PhysRevLett.95.127207,PhysRevLett.95.127206},
the geometric frustration can stabilize the SS states.
However, the previously studied systems have no tunneling frustration, and therefore
can be reduced to non-frustrated models \cite{PhysRevLett.102.017203}.
Only a few literatures concentrate on the corresponding SS physics with frustrated
tunneling \cite{PhysRevA.85.013642,PhysRevB.93.144508,PhysRevLett.102.017203,PhysRevLett.107.146803},
since the standard quantum Monte-Carlo simulation suffers from the notorious sign problem
which prevents a detailed numerical investigation.

We investigate the EBH model via a tensor network state (TNS) method \cite{vidal03,verstraete04,schuch08},
in which the many-particle wave functions can be represented by
the products of tensors.
The TNS wave functions are unbiased and are systematically improvable by increasing bond dimensions $D$.
The method is free of the sign problem and therefore ideal for the frustrated EBH model.
In this work, we use a special TNS, known as the string bond state (SBS)\cite{schuch08}.
More details of the method can be found in Ref.~\cite{Liu2015} and in Appendix A.
We simulate the EBH model on a 10$\times$10 square lattice with periodic boundary condition.
We use the virtual bond dimension for the tensors
up to $D$=10, which converges the total energies with errors less than 10$^{-2}$ per site
for the phase separation states, and  10$^{-3}$ per site for other phases.

\section{Results}

\subsection{Numerical results in the hardcore limit}

To verify our method, we first reproduce the phase diagram for the case of $t_2$=0,
which has been extensively investigated by QMC \cite{PhysRevLett.74.2527,PhysRevLett.84.1599,PhysRevLett.94.207202}.
Our results are in good agreement with the previous QMC results (See Appendix B).
Particularly, both QMC and our results show that
in the hardcore limit, doping $\rho=1/2$ with additional particles or holes
leads to phase separation, which is
a mixture of SF and checkerboard crystal (CBC), instead of a SS state.

\begin{figure}[htpb]
  \centering
  \includegraphics[width=.7\linewidth]{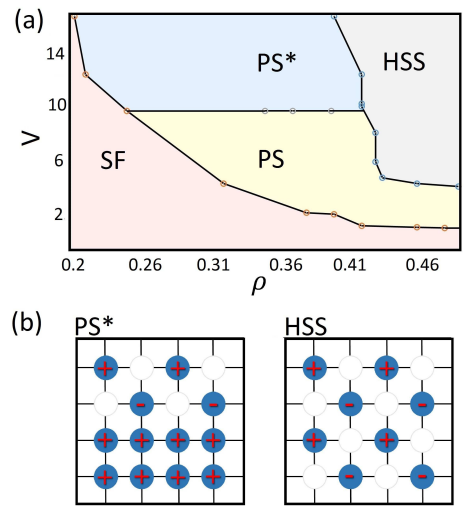}
  \caption{(Color online) (a) The ground state phase diagram
  of the frustrated EBH model
  in the hardcore limit with $t_2$=-0.3. Four phases have been identified including
  a superfluid (SF) phase, a half-supersolid (HSS) phase and phase separation (PS)
  consists SS and SF, and PS* consists of SF and HSS. (b) The schematic plot of the on-site
  $\langle n_i \rangle$ for the HSS and PS* state.
  The empty circles represent $\langle n_i \rangle$=0. The ``$\pm$'' signs in the blue circles indicate the
  on-site boson phase $\arg [\langle b_i \rangle]$.
  }
  \label{t03}
\end{figure}

We then apply the TNS method to study the model with $t_2<$0.
Figure~\ref{t03} shows a typical phase diagram of hardcore boson with
frustrated $t_2=-0.3$
in the $\rho-V$ plane for $\rho \leq$ 0.5. The phase diagram has a mirror symmetry
about $\rho =$ 0.5, because of the particle-hole symmetry.
The most drastic change of the phase diagram
is that a nontrivial half SS (HSS) phase with the checkerboard pattern
emerges at $\rho \in (0.4,0.5)$ and $V>3$,
compared to the phase separation regime in the $t_2$=0 case.
The presence of $t_2$ provides another
hopping channel to lower the energy of
the system for doped holes (particles) within the occupied (unoccupied)
sublattice. Since $t_2$ hopping only happens within one sublattice, the checkerboard pattern is preserved.
This is very different from the $t_2$=0 case,  where doped holes
form a planar domain wall due to NN hopping $t_1$, which
destroys the uniform CBC background and results in phase separation.

To illustrate the nontrivial properties of such HSS phase, we schematically show the on-site boson density
$\langle n_i \rangle$= $\langle b_i^{\dag}b_i \rangle$ of the
HSS phase in Fig. \ref{t03}(b), in which $\avg(n_i)$ forms two sublattices
(the detailed numerical results are given in Appendix B).
Interestingly, in one of the sublattices
both $\avg(n_i)$ and $\avg(\delta n_i)$ are uniform distributed, which corresponds to an uniform SF phase,
and the other sublattice is either fully unoccupied ($\rho<1/2$) or occupied ($\rho\ge 1/2$) with one particle per site, which is precisely a MI.
Since the two interlaced sublattices result in a checkerboard-type density distribution, the whole system possesses both crystalline and SF orders, and can be viewed as some kind of SS phase. However, the simultaneous existence of SF and MI phases in different sublattices makes it
very different from traditional checkerboard SS in the lattice, in which both sublattices have SF order but with different densities.
Since SF order exists only within one sublattice, we call this phase as HSS.

The HSS phases are further distinguished from usual SS phases by
their non-trivial phase patterns $\arg[\langle b_{i} \rangle]$, 
as shown in Fig. 1(b). The HSS phase is composed of the equal superposition of two strip orders with $\vec{k}=(\pi,0)$ and $(0,\pi)$.
For each site $i$ in the MI sublattice,
the summation $\sum_{\overline{i}} \langle b_{\overline{i}} \rangle$ over all NN sites within the SF sublattice vanishes
due to the phase patterns shown in Fig.~\ref{t03}(b).
There is no net particle tunneling between the two sublattices even in the presence of the non-vanishing NN hopping $t_1$, and therefore, the MI sublattice is stabilized even away from half-filling.

The PS* phase in Fig.~\ref{t03} consists of both the HSS phase and the SF
phase that are separated in the real space, as schematically shown
in Fig.~\ref{t03}(b).
The phase boundaries between the SF and HSS are determined by the real space distribution of
$\langle n_i\rangle$ and  $\langle \delta n_i\rangle$ shown in Appendix B.
In the SF region, both $\langle n_i \rangle$ and $\langle \delta n_i \rangle$
distribute almost uniformly, in sharp contrast with the HSS region.
The usual PS (CBC+SF) appears only when $V$ is small.
In addition, there exists a direct first-order transition from PS to HSS. This is somehow different from the
mean-field treatment, where the PS* is always sandwiched between them as shown in Appendix C.
Within numerical precisions, the critical point ($V_c=10$) between PS and PS*
is almost independent of the filling
factor $\rho$, which is consistent with the mean-field predictions discussed in Appendix C.

\begin{figure}
  \centering
  \includegraphics[width=.75\linewidth]{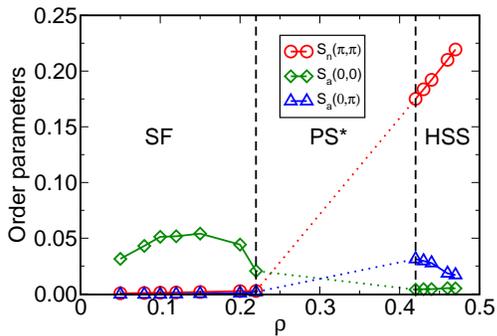}
  \caption{(Color online) Order parameters $S_n(\pi,\pi)$, $S_a(0,0)$ and $S_a(0,\pi)$ as functions of the filling factor $\rho$
   at $V$=12 and $t_2$=-0.3 in the hardcore limit.
  The system shows SF to PS* and PS* to HSS phase transitions as $\rho$ increase from 0 to 0.5.}
    \label{hardcore-op}
\end{figure}

To further characterize the properties of the different phases,
we calculate the static structure factor at $\vec{k}=(\pi,\pi)$ as
\bea
S_n(\vec{k})=\frac{1}{M}\sum_{i,j}\langle n_in_j \rangle e^{-i\vec{k}\cdot(\vec{r}_i-\vec{r}_j)},
\eea
which can be used to identify the
diagonal long-range order in the system. Here $M$ is the total number of lattice sites.
The SF orders are measured by the correlation function defined by
\bea
S_a(\vec{k})=\frac{1}{M}\sum_{i,j}\langle b_i^\dag b_j \rangle e^{-i\vec{k}\cdot(\vec{r}_i-\vec{r}_j)}.
\eea
While generally the SF phase is characterize by the sharp peak of $S_a(\vec{k})$
around $\vec{k}$=(0,0), SS phase is identified by the presence of both $S_a[\vec{k}=(0,0)]$
and $S_a[\vec{k}=(\pi,\pi)]$ accompanied by a nonzero
structure factor $S_n[\vec{k}=(\pi,\pi)]$, which indicates a crystal order.
The HSS phase has equal $S_a(\vec{k})$ around
$\vec{k}=(\pi,0)$ and $(0,\pi)$ along with the presence of the long-range
density wave order as explained in previous text.
Especially, when $\rho>1/2$, the occupied MI sublattice induces a
strong background signal to all $S_a(\vec{k})$,
which clearly discriminates such peculiar HSS states from the usual checkerboard SS phase.

Figure~\ref{hardcore-op} displays the calculated order parameters $S_n(\vec{k})$ and $S_a(\vec{k})$
of the ground state as a function of $\rho$ for fixed $V=12$ and $t_2$=-0.3.
When the filling factor $\rho$ is small, the ground state is a SF characterized by nonzero $S_a(0,0)$ and vanishing $S_n(\pi,\pi)$.
Increasing $\rho$ across a critical value ($\rho \approx$0.2), SF is unstable towards an intermediate PS* region composed of HSS and SF.
For $\rho\geq 0.4$, the SF component disappears, and the system becomes a uniform HSS state with nonzero $S_n(\pi,\pi)$ and $S_a(\pi,0)$ [and $S_a(0,\pi)$].
Approaching half filling, all the order parameters $S_a(\pi,0)$ [and $S_a(0,\pi)$] vanish except $S_n(\pi,\pi)$, which corresponds to the CBC phase.

\begin{figure}[t]
    \centering
    \includegraphics[width=.65\linewidth]{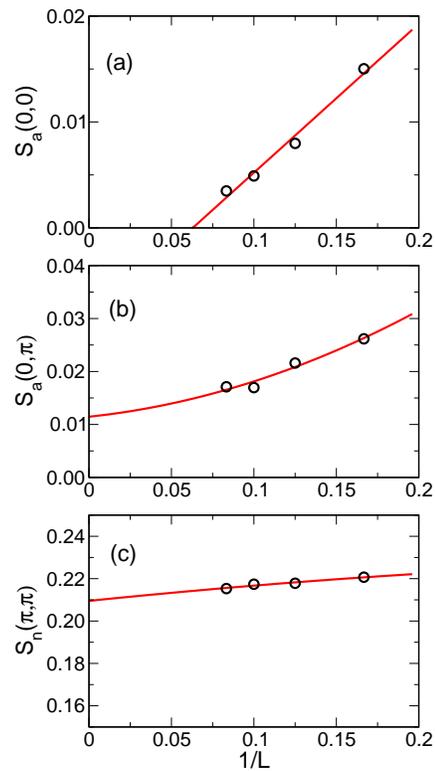}
    \caption{(Color online) The order parameters (a)$S_a(0,0)$, (b) $S_a(0,\pi)$, and (c) $S_n(\pi,\pi)$ as functions of $1/L$, for $L$=6,~8,~10,~12, with $t_2$=-0.3,
    $V$=12, and $\rho \approx$ 0.46. As  $L \rightarrow \infty$, $S_a(0,0)$=0, whereas both $S_a(0,\pi)$ and $S_n(\pi,\pi)$ are finite, suggest that
    the HSS phase is stable in the thermodynamic limit. }
    \label{FiniteSize}
\end{figure}

To see if the HSS phase
is stable in the thermodynamic limit, we calculate
the order parameters $S_a(0,0)$, $S_a(0,\pi)$ and $S_n(\pi,\pi)$ with $t_2$=-0.3, $V$=12, on the $L \times L$ lattices, where $L$=6,~8,~10,~12,
at $\rho \approx$ 0.46, and perform finite size scaling. The results are shown in Fig.~\ref{FiniteSize}(a), (b), (c)
respectively. As $L \rightarrow \infty$, we have $S_a(0,0)$=0, whereas both $S_a(0,\pi)$ and $S_n(\pi,\pi)$ are finite.
These results suggest that the HSS phase is indeed stable in the thermodynamic limit.

\begin{figure}
  \centering
  \includegraphics[width=0.75\linewidth]{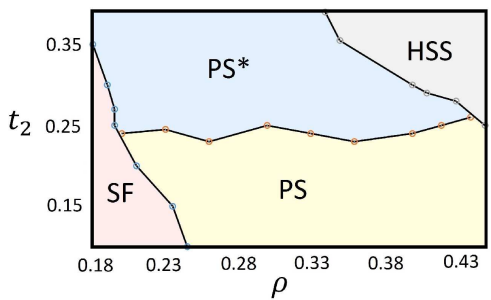}
  \caption{(Color online) The ground state phase diagram of the hardcore frustrated EBH model
   in the $t_2$-$\rho$ plane, with $V$=16. }
  \label{t2}
\end{figure}

To show how the presence of frustration, $t_2$, affect these phases, we also plot the phase diagram in
the $t_2-\rho$ plane for fixed $V=16$, as shown in Fig. \ref{t2}. For small $t_2$, PS appears immediately as
we dope the crystal at $\rho=1/2$ with holes. Increasing $t_2$ makes it more energetically
favorable for the additional holes hopping within the occupied sublattice to form the uniform HSS.
Meanwhile the usual PS is replaced with PS* when $|t_2|\geq 0.24$.
The calculation indicates that HSS should dominate the phase diagram while the PS* region shrinks as we approach the
critical point at $|t_2|=t_1/2$.
This is very different from the non-frustrated case, where stable SS is missing for $\rho<1/2$ even for the soft-core case.
Therefore introducing the frustration hopping term is a
promising route to observe the (H)SS phase.

\subsection{SS phases in the softcore case}

When going beyond the hardcore limit, the EBH model shows far more rich phase diagrams depending on the values of the on-site interaction strength $U$.
In the absence of NNN hopping, previous results show that doping $\rho=1/2$ with additional particles may lead to
a stable SS phase\cite{PhysRevLett.94.207202}.
The key issue here is how this general picture changes due to the presence of HSS phase for $t_2 \neq 0$.

\begin{figure}
  \centering
  \includegraphics[width=0.75\linewidth]{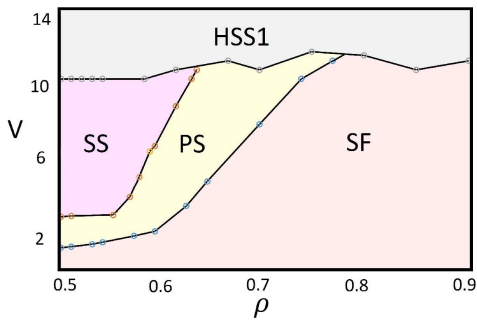}
  \caption{(Color online) The ground state phase diagram of the softcore frustrated
  EBH model in the $\rho$-$V$ plane with $t_2=-0.3$ and $U=40$.}
  \label{U40t03}
\end{figure}

Intuitively, when $2\Delta=zV-U \sim 0$, the doped particles can be placed on either of the two sublattices due to the vanishing energy difference. Therefore, particles can hop around the whole lattice due to NN hopping, which results in the usual checkerboard SS. When  $\Delta < 0$, it is more energetically favorable for the additional particles to occupy the empty sublattice, while the other one remains to be a MI. This indicates that the HSS state discussed above should be more robust for large $U$, and is consistent with those discussed in the hardcore limit. For $\Delta > 0$, the additional particles tend to locate on the occupied lattice site while the empty sublattice remains unchanged. Therefore, there should also exist another kind of HSS phase (HSS1) in this limit
(see Appendix C for details).

Figure \ref{U40t03} depicts the phase diagram
with a constant on-site interaction $U$=40 and $t_2$=-0.3.
The results for $\rho<$0.5 is almost the same as the hardcore limit,
and therefore is not shown.
For these parameters, a stable conventional SS does exist whereas the HSS phase
completely disappears above the half-filling.
The PS* phase consisting of HSS and SF is also missing for parameters we used here.
However, the region of SS phase is greatly shrunk due to $t_2\neq0$
compared with usual extended BH model.
Moreover, the PS regime extends up to the critical line $zV=U$, and sandwiched between SS and SF phases.
When filling factor $\rho \rightarrow 1$, the system is a SF state below the critical line
$zV=U$.
Above $zV>U$, a new HSS1 phase dominates
the phase diagram for all $\rho$.
HSS1 is different from the HSS state in the $\rho>$0.5 region which
has particle-hole symmetry to the HSS state in the $\rho<$0.5 region, i.e.,
it has one sublattice that is fully occupied by one boson per site, without
fluctuation, and the other sublattice is SF.
The HSS1 also have two sublattices with one sublattice being unoccupied, and
the other one being SF with the occupation number larger than one per site.
Interestingly, there is a MI transition in one sublattice
going from the SS phase to the HSS or the HSS1 phases.

\begin{figure}
  \centering
  \includegraphics[width=0.75\linewidth]{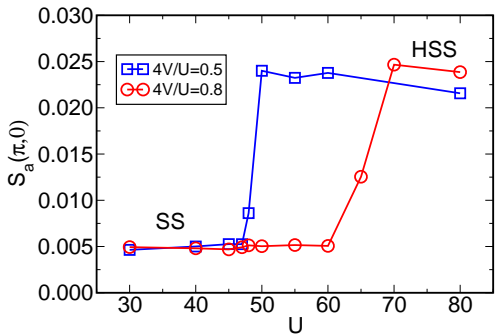}
  \caption{(Color online) The order parameters $S_a(\pi,0)$ as functions of $U$ at $t_2$=-0.3, $\rho$=0.55, for different ratios $zV/U=0.8$ and $0.5$ respectively. }
  \label{softcore-op}
\end{figure}

To see how the HSS phase that presented in the hardcore limit disappears
at $U$=40,  we show in
Fig. \ref{softcore-op} the order parameters $S_a(\pi,0)$ as functions of $U$ for different ratios
$zV/U$=0.8 and 0.5 respectively at $\rho$=0.55. At large $U$, the HSS phase does exist,
but it undergoes a transition to the SS state as $U$ decreases.
The transition from HSS to SS is located at $U$$\sim$50 for $zV/U$=0.5 and $U$$\sim$65 for $zV/U$=0.8.
The SS and HSS phases can coexist in the phase diagram for $U \in$(50, 60).
These results suggest that for a given $U$, the region with larger $V$ will become SS state first,
whereas the region with smaller $V$ can still be HSS.
When $U$=40, the SS-HSS transition may happen at a rather small $V$.
However, in this case, a PS phase would be more
stable, and therefore no HSS phase appear in Fig.~\ref{U40t03}.

\begin{table}[h]
    \caption{ Characteristic order parameters for different supersolid phases discussed in the paper, 
    where $\vec{k}_0=(0,0)$, $\vec{k}_1=(\pi,0)$, $\vec{k}_2=(0,\pi)$, and $\vec{k}_3=(\pi,\pi)$. $S'_a$ describes the background noise for all $\vec{k}$  due to the presence of Mott-insulator sublattice above half-filling. }
    \label{table:s1}
    \begin{center}
    \begin{tabular}{cccc}
        \hline\hline
           & HSS ($\rho<\frac{1}{2}$) or HSS1 & HSS ($\rho>\frac{1}{2}$) & SS  \\
        \hline
         \multirow{2}{*}{$S_a(\vec{k})$}  & \multirow{2}{*}{$S_{a,\vec{k}_1}=S_{a,\vec{k}_2}\neq 0$} &
         $S_{a,\vec{k}_1}=S_{a,\vec{k}_2}\neq 0$, & $S_{a,\vec{k}_0} \neq 0$ \\
         & & and $S'_a=\frac{1}{2}$ & $S_{a,\vec{k}_3} \neq 0$  \\
        \hline
         \multirow{2}{*}{$S_n(\vec{k})$} & \multirow{2}{*}{$S_{n,\vec{k}_0}=S_{n,\vec{k}_3}\neq 0$} & $S_{n,\vec{k}_0} \neq 0$ & $S_{n,\vec{k}_0} \neq 0$ \\
         & & $S_{n,\vec{k}_3} \neq 0$ & $S_{n,\vec{k}_3} \neq 0$  \\
        \hline
    \end{tabular}
    \end{center}
\end{table}

\section{Discussion}

Experimentally, frustrated NNN hopping can be realized using spin-dependent lattice via Raman process,
as demonstrated in Ref.~\onlinecite{PhysRevA.81.021602}.
The effective NN repulsion can be implemented using dipolar bosons \cite{griesmaier2005bose,lu2011strongly,aikawa2012bose} or molecules with tunable ratio $V/U$ via Feshbach resonance.
Finally, the predicted phases can be identified from their typical density and momentum distributions shown in Table I, which can be detected using the time-of-flight (TOF) imaging method combined with the in situ detections.

\section{Summary}

We have studied the phase diagram of the extended Bose-Hubbard models with frustrated hoppings via TNS method. We show that a peculiar HSS phase can be stabilized below half filling even in the hardcore limit, which is absent from the usual non-frustrated Bose-Hubbard models. For softcore model, the competition of on-site interaction and frustration leads to different types SS phases above half filling.
The frustrated models proposed here can be implemented in current experimental systems with no fundamental challenge.
In this work we focus on the case $|t_2|<t_1/2$, but it would be interesting to see how these predicted phases change when approaching the limit $|t_2|\sim t_1/2$, where a spin liquid state is predicted to exist.

\section*{Acknowledgement}
This work was funded by the Chinese National Science Foundation (Grant number 11374275,11574294,11474266,11474267,91536219),
the National Key Research and Development Program of China (Grants No. 2016YFB0201202),
and the Strategic Priority Research Program (B)
of the Chinese Academy of Sciences (Grant No. XDB01030200).
The numerical calculations have been done on the USTC HPC facilities.

\appendix

\section{THE STRING BOND STATE METHOD}
\label{sec:A1}

\begin{figure}[htpb]
    \centering
    \includegraphics[width=.45\linewidth]{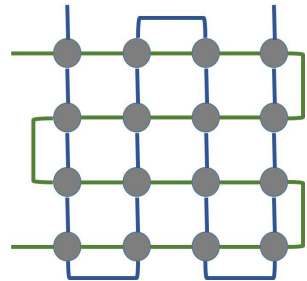}
    \caption{The string-bond states (SBS) pattern used in the calculations,
      which is made up of two long strings.}
    \label{SBS}
\end{figure}

Recently developed tensor network states (TNS) methods, including the
matrix product states (MPS), \cite{vidal03, vidal04,
fannes92} the projected entangled pair states (PEPS),
\cite{verstraete04} string bond states(SBS)~\cite{schuch08}, and the multi-scale entanglement
renormalization ansatz (MERA)~\cite{vidal07}etc. provide a promising scheme to solve
the long-standing quantum many-body problems.
In this scheme, the variational space can be
represented by polynomially scaled parameters, instead of exponential ones.
Once we have the TNS representation of the many-particle wave functions, the
ground state energies, as well as corresponding wave functions, can be obtained
variationally.

In this work, we investigate the ground state properties of the extended Bose-Hubbard model
using SBS.~\cite{schuch08}
The wave functions represented in SBS form can be written as,
\begin{equation}
  | \Psi \rangle = \sum_{s_1 \cdots s_N}^d \prod_{p \in
    \mathcal{P}}{\rm{Tr}}\Big{[} \prod_{k \in p}A_{p,k}^{s_k} \Big{]}
  \Big{|}s_1 \cdots s_N \Big{\rangle},
\end{equation}
where $\mathcal {P}$ is a certain string pattern which contains a set of
strings $p$. The product of matrices $A_{p,k}^{s_k}$ with
virtual bond dimension $D$
over $k \in p$ means over the sites $k$ in the order appearing in the
string $p$, and $d$ is the dimension of the physical indices $s_k$
on site $k$. In this case, $d$ is the possible boson occupation number at each site.
We use $d$=2 for the hardcore boson, and $d$=4
in the soft core case (i.e., the maximum allowed boson on each site is 3).
We use two long strings
as shown in Fig.~\ref{SBS}.
This type of SBS satisfies area law,~\cite{schuch08,wangzhen13}
and the results can be systematically converged
by increasing the bond dimension $D$.

For a given a Hamiltonian $H$,
the total energy of the system is a function of the tensors at each lattice
site $A_{p,k}^{s_k}$, i.e., $E=E(\{A_{p,k}^{s_k}\})$.
We recently developed an efficient algorithm to obtain
the ground state wave function and corresponding energy,
by mapping the problem to optimizing the total energy of
a classical mechanical system, in which
the elements  $a_{ij}^{s_k}(p,k)$ of the tensor ${A_{p,k}^{s_k}}$ are treated
as the generalized coordinates of the system.
The ground state wave function and total energy is then obtained
via a replica-exchange molecular dynamic simulation.
Details of the method are presented in Ref.~\onlinecite{Liu2015}.
The replica exchange (also known as parallel tempering)~\cite{swendsen86, geyer_book} MD
method can effectively avoid the system being
trapped in local minima, which is very important for
accurate simulation of complex states such as
phase separations.

In our simulations, we use $M$=48 temperatures. Initially, the temperatures
distribute exponentially between the highest (1/$\beta_0$=0.01)
and lowest (1/$\beta_{M-1}$=10$^{-6}$)  temperatures. For each temperature,
we start from random tensors.
During the simulations, we adjust the temperatures after configuration
exchange for 10 times, whereas there are 300 MD steps between the two
configuration exchanges,
with a step length $\Delta t$=0.01.
For each MD step, we sample about 40000
configurations. The energies used for temperature exchange
are averaged over 250 MD steps.
After we finish the replica-exchange MD optimization, we further decrease the
temperature to 10$^{-7}$ to obtain more accurate results.

\begin{figure}[h]
    \centering
    \includegraphics[width=.90\linewidth]{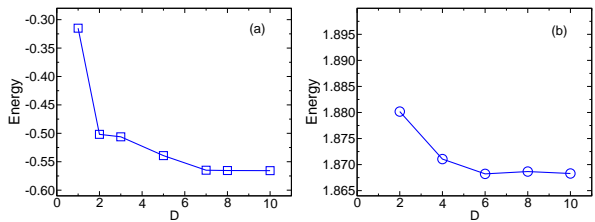}
    \caption{(color online) The total energy per site as functions of bond dimension $D$
    for (a) the hardcore limit, with $V$=1, $t_2$=-0.3, $\rho$=0.32, which is
     in the SF phase; and (b) the softcore case with $U$=40, $V$=12, $t_2$=-0.3 and $\rho$=0.55,
     which is in the HSS1 phase.}
    \label{ED}
\end{figure}

Figure~\ref{ED}(a),(b) shows the convergence of total energy
per site with increasing bond dimension $D$ in for two typical parameter sets.
When $D$=10, the total energies converge to less than 10$^{-2}$ per site
for the phase separation states, and 10$^{-3}$ per site for pure phases.

\section{Some detailed numerical results}

\subsection{Results for $t_2$=0 model}

\begin{figure}[h]
    \centering
    \includegraphics[width=.65\linewidth]{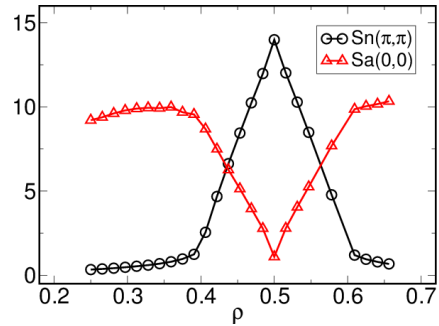}
    \caption{(Color online) The structure factors $S_n(\pi,\pi)$ and $S_a(0,0)$ as functions of density $\rho$ for the $t_2=0$, $V=3$ hardcore
    Bose-Hubbard model on a 8$\times$8 lattice.}
    \label{hard}
\end{figure}

\begin{figure}[h]
    \centering
    \includegraphics[width=.75\linewidth]{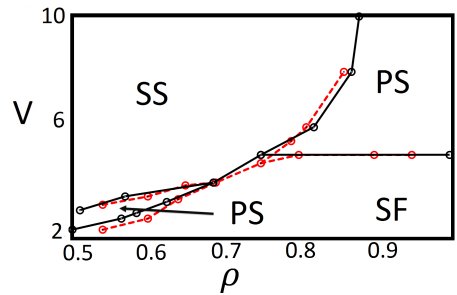}
    \caption{(Color online) The phase diagrams of the extended Bose-Hubbard model with $U=20$, $t_2=0$ on 10$\times$10 lattice in the $V$-$\rho$ plane.
    We show only the region of $\rho>0.5$. The red dashed lines are calculated by SBS method, whereas the black solid lines
    are extracted from Ref.~\onlinecite{PhysRevLett.94.207202}}.
    \label{soft}
\end{figure}

To benchmark our method, we first calculate the phase diagrams for $t_2$=0 model, which has no tunneling frustration, and compare the results
to those obtained from quantum Monte Carlo (QMC) simulations for the hardcore bosons~\cite{PhysRevLett.84.1599}
and soft-core bosons~\cite{PhysRevLett.94.207202}.
In Fig.~\ref{hard}, we plot the structure factors $S_n(\pi,\pi)$ and $S_a(0,0)$ as functions of filling density $\rho$ for the hardcore model.
The results are obtained on a 8$\times$8 lattice with $V=3$. The structure factors and phase boundaries
obtained using the SBS are in good agreement with those obtained from QMC with the same parameters (see Fig. 1 of Ref.~\onlinecite{PhysRevLett.84.1599}).
In Fig.~\ref{soft}, we compare the $V$-$\rho$ phases diagrams for $\rho \ge$0.5 with $t_2$=0, $U=20$ on a 10$\times$10 lattice
 calculated by the SBS method (red dashed lines) with those obtained
from QMC method with the same parameters (black solid lines extracted from Ref.~\onlinecite{PhysRevLett.94.207202}), which have good agreement.

\begin{figure*}[tb]
    \centering
   \includegraphics[width=0.75\linewidth]{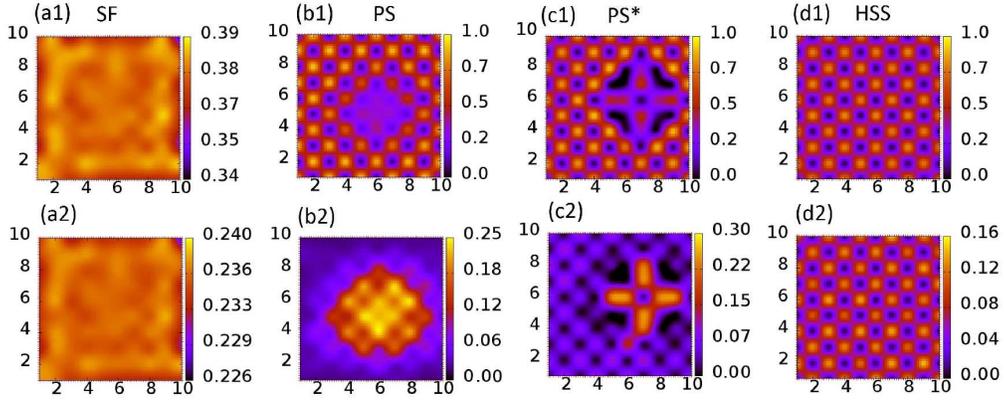}
    \caption{ (Color online) Upper panel: Typical boson density distribution $\avg(n_i)$=$\avg(b_i^{\dag}b_i)$ for (a1) superfluid
      (SF), (b1) phase separation (PS), (c1) PS* and (d1) half-supersolid
      (HSS) phases in the hardcore limit and filling $\rho <$0.5.
              Lower panel: (a2 - d2) Corresponding density fluctuation
      $\avg(\delta n_i)$ for each phase.}
    \label{hardcore}
\end{figure*}

\begin{figure*}[tb]
    \centering
    \includegraphics[width=.75\linewidth]{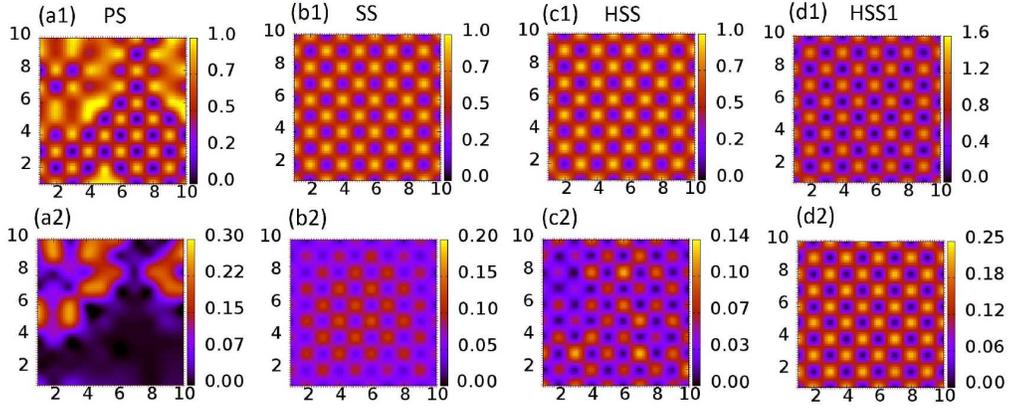}
    \caption{ (Color oneline) Upper panel: Typical boson density distribution $\avg(n_i)$
      for (a1) PS, (b1) SS, (c1) HSS and (d1) HSS1 phases
      in the softcore case and filling $\rho >$0.5.
        Lower panel: (a2 - d2) Corresponding density fluctuation
      $\avg(\delta n_i)$ for each phase.}
    \label{softcore}
\end{figure*}

\subsection{The density distribution and fluctuation of different phases}

The various phases emerge from the frustrated EBH model can
be distinguished from their density distributions $\avg(n_i)$=$\avg(b_i^{\dag}b_i)$ and
density fluctuations  $\avg(\delta n_i)$ in real space.
We show in Fig.~\ref{hardcore} the typical on-site density distribution $\avg(n_i)$
(upper panel) and corresponding density fluctuation $\avg(\delta n_i)$ (lower
panel) for the frustrated Bose-Hubbard model in the hardcore limit and
filling $\rho <$ 0.5,
including the superfluid (SF), half-supersolid (HSS), phase separation (PS)  and PS* phases.
The SF phase have almost uniformly distributed density $\avg(n_i)$ and so does the density fluctuation
$\avg(\delta n_i)$ on each site.
The HSS phase clearly shows two sublattices, with one sublattice is fully unoccupied
and absent of $\avg(\delta n_i)$. The other sublattice is SS, with non-vanish
boson density and density fluctuation.
The PS phase shows distinct regions in real space with one being SF, and the
other being CBS, whereas in the PS* phase,
there are also two distinct regions with one being SF, and the other being HSS.

Figure~\ref{softcore} shows the typical density distribution $\avg(n_i)$
(upper panel) and corresponding density fluctuation $\avg(\delta n_i)$ (lower
panel) for the frustrated EBH model in the softcore case and
filling $\rho >$ 0.5, including PS, SS, HSS  and HSS1 phases.
The PS phase is composed of the SF and CBS phase.
The SS phase clearly has both solid order and SF order.
It has two sublattices, both have non-vanish density distribution $\avg(n_i)$
and density fluctuation $\avg(\delta n_i)$ on each site
which is different from those of the HSS phase.
In the softcore case with  $\rho >$ 0.5, there are two HSS phase, namely HSS
and HSS1. When $\rho >$ 0.5, one of the sublattice of HSS is fully occupied,
with occupation number equal one. The other sublattice is SF. This phase
can be viewed as a particle-hole mirror image of the HSS phase for $\rho <$ 0.5.
The HSS1 phase also has two sublattices, but it
is different from the HSS phase, as it has one fully
unoccupied sublattice, and the other one being SS with the
occupation number larger than one.

\section{Results from Mean field theory}

Since the standard mean-field calculation cannot capture the main physics related to phase separation, we employ the method introduced by Kimura \cite{kimura2011gutzwiller} which combines the Gutzwiller variational wavefunction and linear programming method. It has been proved that such method can effectively describe the phase diagram within a canonical ensemble with fixed total boson number, especially in the hardcore limit. Specifically, we assume the total free energy of the system can be written as the combination of two different phases $E_{t}=\gamma E_A+(1-\gamma) E_B$ with $\gamma$ the area ratio of phase A in the entire system. To obtain the ground state, we minimize $E_t$ subject to the particle-number condition as $N_t=\gamma N_A+(1-\gamma)N_B$. When $\gamma=0$ or $1$, the system stays in a uniform phase B or A without phase separation. While for $\gamma \in (0,1)$, a mixture of phases A and B occurs which indicates the existence of phase separation.

In the case of $t_2=0$, the corresponding phase diagram has been widely studied using standard quantum Monte-Carlo method. The calculation shows that when $U/|t_1| \rightarrow \infty$, the supersolid state is unstable due to the presence of PS into a mixture of uniform SF and crystal.
This is because away from half filling, the additional holes for $\rho<$ 0.5 tend to
to form a planar domain wall with the energy gain $\Delta E_1 \sim -c t_1$ per holes. While for an isolated hole, the kinetic energy gain is only estimated as $\Delta E'_1 \sim -4 t_1^2/3V$. This leads to the instability of the crystal phase at $\rho=1/2$ when doped with holes, which finally develops into the mixture a uniform superfluid and the checkerboard solid. In the other limit with $t_1=0$, the presence of NNN hopping $t_2$ provides another channel to low the energy of the system away from half filling. The additional holes can hop around the background of a checkerboard density-wave with the energy gain determined by $\Delta E_2 \sim -4 t_2$. In this case, the superfluid can exist for $\rho<1/2$ accompany with a checkerboard density-wave, which is the characteristic feature of supersolid (SS) phases. Since we have $t_2<0$, the interference effect ensures that particles only reside within one sublattice. Such half-SS (HSS) state is different from the usual SS state when $t_2>0$, where additional holes can jump between the two sublattices.

In the presence of both $t_1$ and $t_2$, the competition of above two mechanisms leads to a rich phase diagram in the $\rho-V$ plane. The system can be in a HSS state or PS depending closely on the relative strength of
$t_1$, $t_2$, and $V$ respectively. The total energy gain of PS can be estimated as $\Delta E_1 \sim -c t_1 + d V$ taking into account the interaction between neighboring lattice sites. When both $t_2$ and $V$ are comparatively small, the PS state has much lower energy which excludes the possibility of SS state away from half filling. However, in the opposite case, we have $\Delta E_2 < \Delta E_1$, which indicates that the SS phase is stabilized by the NNN hopping and NN repulsive interaction.

To obtain the boundaries of various phases, we employ a perturbation analysis using the canonical ensemble.

\subsection{phase separation into SF and HSS state from HSS}

For relatively larger $V$, the presence of NNN hopping $t_2$ can stabilize the HSS phase away from half filling, where SF exists only within one sublattice.  While in the opposite limit with $\rho \sim 0$ or $(1-\rho) \sim 0$, a uniform SF state has much low energy. It is expected that in the intermediate regime, there exists a PS* state composed of the mixture of SF state and HSS state. This is very different from the normal case with $t_2=0$, where PS occurs immediately when the system is doped with additional holes or particles in the background of CB solid at half filling.

In the hardcore limit, we assume the Gutzwiller wavefunctions of HSS and SF phases to be
\bea
|\psi_{hss}\rangle = d_0 |0\rangle + d_1 |1\rangle, \mbox{ and } |\psi_{sf}\rangle = c_0 |0\rangle + c_1 |1\rangle.  \nn
\eea
The total number of particles per site can be expressed as
\bea
\rho=(1-\gamma_{sf})(\rho-\delta \rho)-\gamma_{sf} \rho_{sf},
\eea
where we have set $\gamma_{sf}$ the ratio of SF phase. The coefficients of wavefunctions $|\psi_{hss}\rangle$ and $|\psi_{sf}\rangle$ are related to mean number of particles through $|d_0|^2=1-|d_1|^2=2(\rho-\delta \rho)$ and $|c_0|^2=1-|c_1|^2=1-\rho_{sf}$. Near the transition boundary, we have $\delta \rho \simeq \gamma_{sf}(\rho_{sf}-\rho)$. The total variational energy of the system is obtained as
\bea
E_t&=&(1-\gamma_{sf})E_{hss}-\gamma_{sf} E_{sf} \nn \\
&\simeq & 2zt_2(\rho-2\rho^2)+\gamma_{sf}  \Big \{  2 [V+2(t_1+t_2)]\rho_{sf}^2  \nn \\
&& \hspace{.8cm}+ 4 (-t_1 - (2-4\rho) t_2)\rho_{sf} + 8 \rho^2 t_2 \Big\}.
\eea
The transition point can then be found by setting the coefficient of $\gamma_{sf}$ to be zero. This is achieved when $\rho_{sf}=4\rho t_2/[4\rho t_2-(t_1+2t_2)]$ for $t_1+2t_2>0$. The corresponding critical values of $V$ and $\rho$ are obtained as
\bea
V_c &=& -2t_2-\frac{(t_1+2t_2)[t_1+2t_2-8t_2\rho(1-\rho)]}{4\rho^2t_2}, \mbox{  and } \nn \\
\rho_c &=& \frac{(t_1+2t_2)}{2\sqrt{|t_2|}(\sqrt{V+2t_1+2t_2}-2\sqrt{|t_2|})}.
\eea
Therefore, for given total number of particles per site, we have a HSS state for $V>V_c$ while PS occurs for $V<V_c$. The same argument applies to $\rho_c$, where PS phase appears when $\rho<\rho_c$ for fixed NN repulsive interaction $V$. In the case of half-filling, we have $V_c=-2(t_1+t_2)-t_1^2/t_2$.

\subsection{PS into HSS and SF phases from SF}

For larger NN repulsive interaction $V$, increase the total particle number leads to the transition from a SF state to the mixture of SF and HSS states. The boundary can also be obtained using similar method discussed above. The constraint of total number of particles per site leads to
\bea
\rho=(1-\gamma)(\rho-\delta \rho) + \gamma \rho_{hss}
\eea
with $\gamma$ and $\rho_{hss}$ corresponding to the ratio and mean number of particles in HSS phase respectively. The total variational energy of the system can be derived  as
\bea
E_t &=& (1-\gamma)E_{sf}+\gamma E_{hss} \nn \\
&=& -z(t_1+t_2)\rho(1-\rho)+\frac{zV}{2}\rho^2  \nn \\
&& + \gamma \frac{z}{2}\Big \{ A\rho^2-4t_2\rho_{hss}^2 -2[A\rho-(t_1+2t_2)] \rho_{hss} \Big \} \nn \\
\eea
with $A=V+2(t_1+t_2)$. Due to the internal symmetry, here we only concentrate on the case with $\rho\in[0,1/2]$. Similar argument applies by replacing $\rho$ with $(1-\rho)$ when $\rho\in(1/2,1]$.

The boundary can be solved by looking at the minimal of coefficient associated with $\gamma$
\bea
f(\rho_{hss})= A\rho^2-4t_2\rho_{hss}^2 -2[A\rho-(t_1+2t_2)] \rho_{hss}.
\eea
Since phase separation occurs only when $f(\rho_{hss})<0$, the critical $V_c$ and $\rho_c$ can be obtained by setting $f(\rho_{hss})=0$  within the regime $\rho\in[0,1/2]$. Generally, it is easy to check that $f(\rho_{hss})$ reaches its minimal
\bea
f(N_{hss})_{min}=A\rho^2 - 4|t_2|\rho_{hss}^2 \nn
\eea
when $\rho_{hss}=-[A\rho-(t_1+2t_2)]/4t_2$, which is less than $1/2$ only when  $V > -2(t_1+t_2)-t_1^2/t_2$. Setting $f(\rho_{hss})_{min}=0$, we obtain $\rho_{hss}=(t_1+2t_2)/2(\sqrt{A|t_2|}-2|t_2|)$. The corresponding critical $\rho$ and $V$ read
\bea
\rho_c &=& \frac{t_1+2t_2}{A-2\sqrt{A|t_2|}},  \mbox{  and  } \nn \\
V_c &=& \frac{1-2\rho}{\rho}(t_1+2t_2)+2\sqrt{t_2(t_2-\frac{t_1+2t_2}{\rho})}.
\eea
When $\rho_{hss}=1/2$, the critical NN repulsive interaction reduces to $\widetilde{V}_c=-2(t_1+t_2)-t_1^2/t_2$ with $\widetilde{\rho}_c=|t_2|/t_1$.
When $V < -2(t_1+t_2)-t_1^2/t_2$, $f(\rho_{hss})$ reach its minimal only when $\rho_{hss}=1/2$. In this case, the boundary is given by
\bea
V_c &=& \frac{\rho^2+(1-\rho)^2}{\rho(1-\rho)} (t_1+t_2), \mbox{  and   }  \nn \\
\rho_c &=& \frac{1}{2}\left(1-\sqrt{\frac{V-2(t_1+t_2)}{V+2(t_1+t_2)}} \right).
\eea

\begin{figure}[htpb]
    \centering
    \includegraphics[width=.960\linewidth]{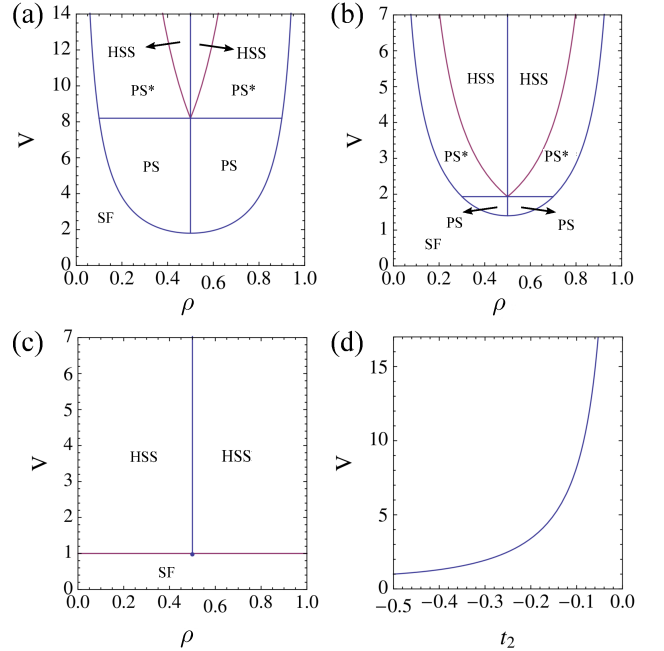}
    \caption{(Color online) Phase diagrams in the hardcore limit in the $\rho-V$ plane for different $t_2=-0.1$ (a), $t_2=-0.3$ (b) , and $t_2=-0.5$ (c) with fixed $t_1=1$. (d) shows the critical $\widetilde{V}_c=-2(t_1+t_2)-t_1^2/t_2$ as a function of $t_1$ and $t_2$. }
    \label{fig:A1}
\end{figure}

The mean-field phase diagrams can then be obtained based on the above analysis, which are depicted in Fig. \ref{fig:A1}.  When $t_2=0$, it has been shown that doping the crystal at $\rho=1/2$ with additional holes or particles leads to the PS between CB solid state and SF state.
The presence of $t_2$ greatly suppresses the regime of this usual PS phase, where HSS phase with additional PS* state between SF and HSS regimes appears on the top of the diagram. The boundaries of the two different PS states is determined by $\widetilde{V}_c=-2(t_1+t_2)-t_1^2/t_2$. Interestingly, at
the mean-field level, such boundary also overlaps with the lower tip of HSS phase at $\rho=1/2$.
We note that in the TNS results, this boundary shifts upwards, as shown in Fig. 1, which
indicates that the mean-filed calculation over-estimates the lower boundary of $\widetilde{V}_c$.
Moreover, HSS states gradually dominate the phase diagram as $|t_2|$ approaches the highly frustrated point $|t_2|=t_1/2$, suggesting that the SS state can be stabilized for arbitrary $\rho<1/2$, which is unattainable in the usual Bose-Hubbard model without frustrated hopping.

\begin{figure}[htpb]
    \centering
    \includegraphics[width=.970\linewidth]{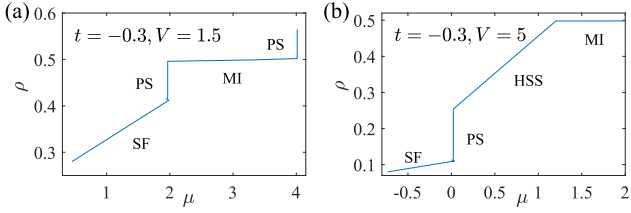}
    \caption{(Color online) $\rho$ versus $\mu$ for fixed $t_2=-0.3$ and different $V=1.5$ [(a)] and $5$ [(b)] respectively in the hardcore limit. The vertical lines show the locations where phase separation occurs, which is similar to the curve obtained from the Maxwell construction for canonical ensembles. }
    \label{fig:A1b}
\end{figure}

 We would like to note that although the above results are obtained within the constraints of fixed total particle numbers, the approximation based on Gutzwiller wavefunction make it different from the usual particle-number-conserved system. In addition, the optimization procedure employed during the calculation also excludes the possibilities of finding uniform phases within the PS regime.  To show this, we also plot the $\rho-\mu$ curves by sweeping $\rho$ across the PS regime in Fig. \ref{fig:A1b}. The appearance of vertical lines indicates that the whole system is the mixture of two different phases, similar to that obtained from the usual Maxwell construction for canonical ensembles.

\subsection{Fate of SS state in the Softcore case}
In the softcore case, the breakdown of particle-hole symmetry around $\rho=1/2$ results in different phases when the additional holes and bosons are introduced. For $\rho<1/2$, the doped holes can hop within the occupied lattices to reduce the kinetic energy, which leads to the HSS phase and is very similar to the hardcore limit $U \rightarrow \infty$. When $\rho>1/2$, the additional bosons can be placed on either sublattice. The system can be in different phases depending on the relative ratio of $t_1$, $t_2$, $U$ and $zV$ respectively.

\begin{figure}[htpb]
    \centering
    \includegraphics[width=.96\linewidth]{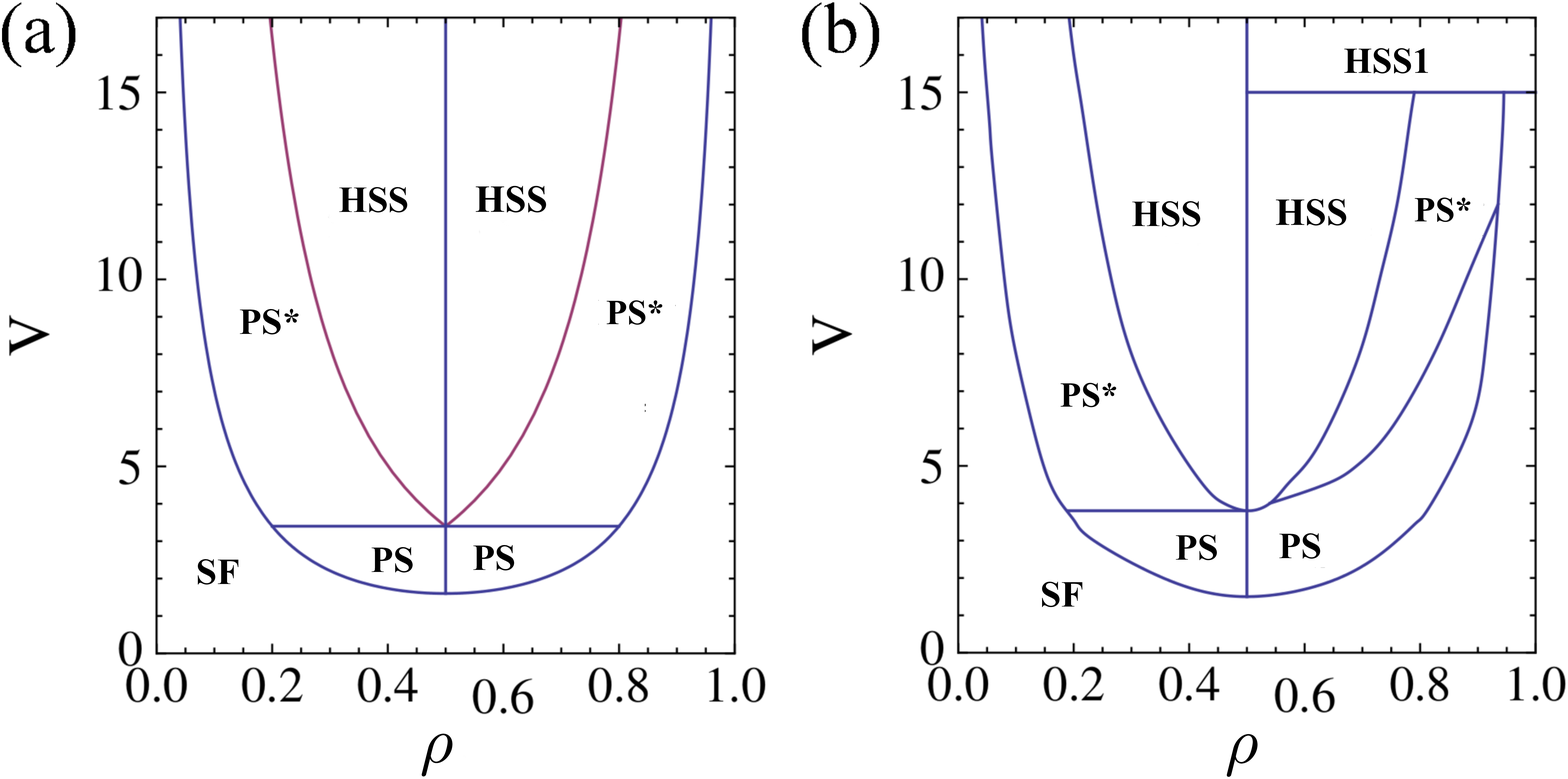}
    \caption{(Color online)  Phase diagrams in the $\rho-V$ plane for $t_2=-0.2$ with different $U=\infty$ (a) and $U=60$ (b). There are two different HSS state for $zV<U$ and $zV>U$ respectively when $\rho>1/2$.  }
    \label{fig:A2}
\end{figure}

Specifically, when $zV<U$, there are two different ways to low the energy of the system. In the first case, the particles are located within the unoccupied sublattice, which can then hop to the NNN site due to the presence of $t_2$. This is a HSS state where superfluid exists only within one sublattice. The total energy gain per particle can be estimated as $\Delta E'_2 = zV - 4 |t_2|$.  On the other hand, when the bosons can hop between occupied and unoccupied site, the energy gain is given by the lowest eigenvalue of the following coupling matrix as
\bea
\left(
\begin{array}{cc}
U+2 z|t_2| & \sqrt{2}zt_1 \\
\sqrt{2}zt_1 & zV+z |t_2|
\end{array}
\right)
\eea
and reads
\bea
\Delta E_- = U+\Delta+3 z |t_2|/2-\sqrt{2(z t_1)^2 + (z|t_2|/2-\Delta)^2}, \nn \\
\eea
where $2\Delta=zV-U$.
The ground state configuration can then be determined by the difference
\bea
\delta E &=& \Delta E_--\Delta E'_2  \nn \\
&=& -\Delta + 5 z|t_2|/2-\sqrt{2(z t_1)^2 + (z|t_2|/2-\Delta)^2}, \nn \\
\eea
which corresponds to a HSS phase when $\Delta E_->\Delta E'_2$, and a SS phase when $\Delta E_-<\Delta E'_2$.
Many interesting features can be obtained from the above formula and we summarized then as follows:
\begin{enumerate}[1)]
\item It is easy to check that for very small NNN hoppings $|t_2|\simeq 0$, we always have $\delta E =-\Delta -\sqrt{2(z t_1)^2 + \Delta^2} <0$, therefore SS phase favors for all $\Delta$ when we dope the crystal at $\rho=1/2$ with particles;
\item When $|\Delta| \gg \max(|t_1|,|t_2|)$, we find $\delta E \sim 8(|t_2|-2 t_1^2/|\Delta|) >0$, which leads to the stabilization of HSS phase in this case. We note that in the hardcore limit, since $|\Delta|\rightarrow \infty$, such HSS phase is always more stable then SS phase, and consistent with the previous discussion;
\item On the other side with $|\Delta| \sim 0$, we have $\delta E \sim 5 z |t_2|/2-\sqrt{2(z t_1)^2 + (z|t_2|/2)^2}<0$ for all $|t_2|<t_1/2$. The ground state is described by a SS state;
\item The boundary between SS phase and HSS phase can be estimated from $\delta E =0$, which gives $|\Delta_c|=2[(t_1/t_2)^2-3]|t_2|$.
\end{enumerate}

The above arguments can also be extended to $zV>U$ with
\bea
\delta E &=& \Delta E_--\Delta E''_2  \nn \\
&=& \Delta + 5 z|t_2|/2-\sqrt{2(z t_1)^2 + (z|t_2|/2-\Delta)^2}, \nn \\
\eea
where $\Delta>0$ and  $\Delta E''_2=U-4|t_2|$ is the total energy of a single boson delocalized within the occupied sublattice due to $t_2$. In the case $|\Delta| \gg \max(|t_1|,|t_2|)$, we have $\delta E \sim 4(3|t_2|-4 t_1^2/|\Delta|) >0$, which indicates that the HSS state is favored. When $\Delta \sim 0$,  we still have $\delta E <0$. Therefore SS state also exists only within a very smaller regimes around the $zV=U$. This is very different from the non-frustrated case, where SS state is stabilized almost for all $zV>U$.

Figure \ref{fig:A2}(b) depicts the phase diagram in the $\rho-V$ plane for $t_2=-0.2$ with finite $U=60$. Since $U \gg (|t_1|,|t_2|)$, the phase diagram is only slightly changed on the $\rho<1/2$ side compared with the hardcore case [Fig. 2(a)]. When $\rho>1/2$, there are two HSS states on the different side of the critical line $zV=U$. When $zV>U$, all particles locate within one sublattice to reduce the interaction energy due to the presence of frustrated hoppings. This corresponds to the HSS state (HSS1) where the mean occupation number is larger than one. In the opposite case with $zV<U$, the additional particles take place the unoccupied lattice to form a sublattice SF state, while the remaining sublattice is still a CB solid state. Therefore, the whole system is composed of a uniform mixture of solid and SF states, which is the usual particle-hole counterpart of the HSS state as both SF and DW order parameters are present in the regime $\rho <$1/2. Similarly, the PS state between the HSS phase and SF phase can be viewed as the mixture of the two phases.

\begin{figure}[htpb]
    \centering
    \includegraphics[width=.96\linewidth]{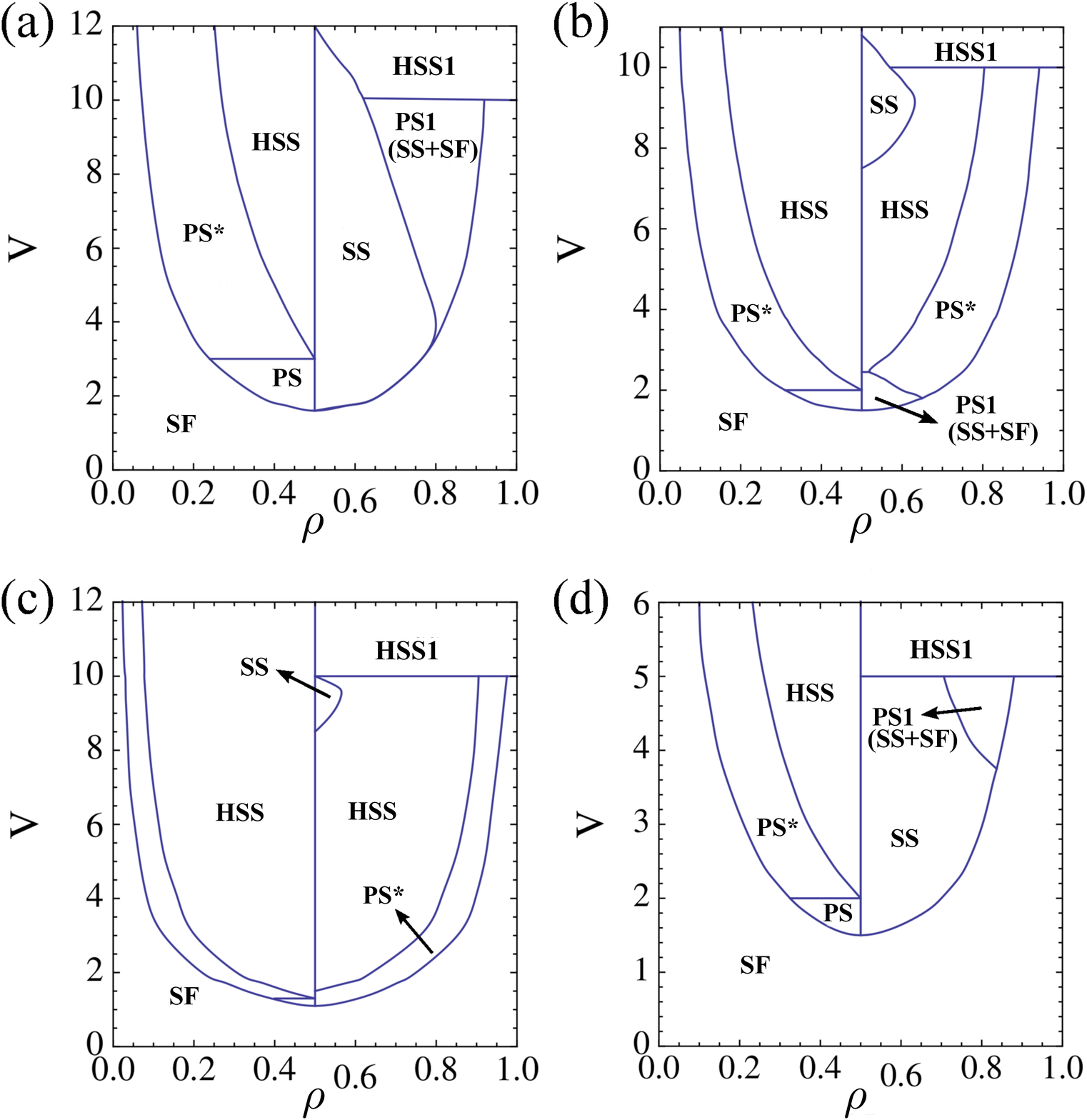}
    \caption{(Color online) Phase diagrams in the $\rho-V$ plane for $U=40$ with different $t_2=-0.2$ (a), $t_2=-0.3$ (b), and $t_4=-0.4$ (c). SS phase becomes more stable as decreasing the on-site interaction $U$. While HSS state can be favored by increasing $|t_2|$, SS exists only within a relatively smaller regime around $zV=U$ by doping additional particles.  (d) shows the phase diagram for $U=20$ and $t_2=-0.2$.   }
    \label{fig:A3}
\end{figure}

Further decreasing $U$ makes it energetically more favorable for additional bosons hopping over both occupied and unoccupied sites.
Such effect is very apparent when $|t_2| \ll t_1/2$, where SS state dominates the phase diagram, as shown in Fig. \ref{fig:A3}. Meanwhile, an intermediated PS occurs between the SS and SF phases.
When $|\Delta| \gg 0$, $\Delta E_-$ can be larger then $\Delta E_1$ as increasing $|t_2|$.
This leads to the stabilization of HSS and HSS1 phases for large $|\Delta|$.
In addition, the regime of SS state shrinks and exists only around the critical point $zV=U$ above half filling.

The above mean-field results have also been checked by our numerical calculation using TNS. Comparatively speaking, the results suggest that HSS phase above half-filling is more fragile than that predicted based on such mean-field treatment. For the same parameters used in Fig. \ref{fig:A3}(b), HSS phase is completely missing when $\rho>1/2$, as shown in Fig. 4 in the main text.  To show the competition between SS and HSS phase as we decrease the onsite interaction $U$, we also plot the transition points along with $U$ for different ratios $zV/U=0.5$, and $0.8$ using TNS, as shown in Fig. 5. The results indicate that along with the decreasing of $U$, SS phase appears first around $zV \sim U$, and dominates the phase diagram as $U$ decreases, which is consistent with the above discussions. We also note that no phase separation regimes composed of the mixture of SS and SF phases are found during the TNS calculation.

%

\end{document}